\documentclass[prb,twocolumn,superscriptaddress,showpacs,floatfix]{revtex4-1}%

\usepackage[english]{babel}
\usepackage[T1]{fontenc}
\usepackage[latin9]{inputenc}
\usepackage{babel}
\usepackage{amsmath}
\usepackage{amssymb}
\usepackage{wasysym}
\usepackage{graphicx}
\usepackage{xcolor}
\usepackage{subfigure}
\usepackage{multirow}
\usepackage{dcolumn}
\usepackage{booktabs}
\usepackage{braket}

\newcommand{\x}{\hspace{0.02cm}~}
\newcommand{\tc}{T$_{\rm C}$}

\newcommand{\tcexp}{T$^{\rm exp}_{\rm C}$}

\definecolor{mydarkgreen}{rgb}{0.0,0.5,0.0}
\newlength\mylen
\begin{document}

\title{Ab initio theory of plasmonic superconductivity within the Eliashberg and density-functional formalisms}

\author{A. Davydov}
\affiliation{Department of Physics, University of Zurich, Winterthurerstrasse 190, 8057 Zurich, Switzerland}
\author{A. Sanna}
\affiliation{Max-Planck-Institut f\"ur Mikrostrukturphysik, Weinberg 2, D-06120 Halle, Germany}
\author{C. Pellegrini}
\affiliation{Fritz Haber Center for Molecular Dynamics, Institute of Chemistry, The Hebrew University of Jerusalem, Jerusalem 91904, Israel}
\author{J.K. Dewhurst}
\affiliation{Max-Planck-Institut f\"ur Mikrostrukturphysik, Weinberg 2, D-06120 Halle, Germany}
\author{S. Sharma}
\affiliation{Max-Born-Institut f\"ur Nichtlineare Optik und Kurzzeitspektroskopie, Max-Born-Strasse 2A, 12489 Berlin, Germany}
\author{E.K.U. Gross}
\affiliation{Fritz Haber Center for Molecular Dynamics, Institute of Chemistry, The Hebrew University of Jerusalem, Jerusalem 91904, Israel}

\definecolor{airforceblue}{rgb}{0.36, 0.54, 0.66}
\definecolor{darkgreen}{rgb}{0.09, 0.45, 0.27}
\definecolor{britishracinggreen}{rgb}{0.0, 0.26,0.15}
\definecolor{deeppink}{rgb}{0.94, 0.19, 0.22}

\begin{abstract}
We extend the two leading methods for the \emph{ab initio} computational description of phonon-mediated superconductors, namely Eliashberg theory and density functional theory for superconductors (SCDFT), to include plasmonic effects. Furthermore, we introduce a hybrid formalism in which the Eliashberg approximation for the electron-phonon coupling is combined with the SCDFT treatment of the dynamically screened Coulomb interaction. The methods have been tested on a set of well-known conventional superconductors by studying how the plasmon contribution affects the phononic mechanism in determining the critical temperature (\tc). Our simulations show that plasmonic SCDFT leads to a good agreement between predicted and measured \tc's, whereas Eliashberg theory considerably overestimates the plasmon-mediated pairing and, therefore, \tc. The hybrid approach, on the other hand, gives results close to SCDFT and overall in excellent agreement with experiments.

\end{abstract}
\date{\today}
\maketitle

\section{Introduction}

Superconductors that do not fit into the standard BCS class, have opened interesting routes for alternative pairing mechanisms, whose applicability is still under debate~\cite{Norman_UnconventionalSC_Science2011,Stewart_UnconventionalSC_AdvPhys2017,Kopelevich_UnstableElusiveSuperconductors_PhysicaC2015}. Developing a first-principles method for the accurate calculation of the critical temperature (\tc ) would not only clarify the microscopic mechanisms of superconductivity, but also contribute to the search for higher temperature superconductors.
It has long been suggested that the key to high temperature superconductivity might be a purely electronic mechanism, that directly exploits the Coulomb repulsion between the electrons to provide their pairing. Many investigations have addressed the role of paramagnetic spin fluctuations in iron-based~\cite{Mazin_spmLaFeAsO_PRL2008,Hirschfeld_GapSumAndStructureFeSC_2011,Mazin_IronBoost_Nature2010} and copper-oxide high temperature superconductors~\cite{Moriya_AFspinfluctuations_RPP2003,Stewart_UnconventionalSC_AdvPhys2017}. Other proposals, instead, have focused on the effective attraction~\cite{KohnLuttinger_NewMechanismForSuperconductivity_PRL1965} appearing in the dynamically screened Coulomb interaction due to the exchange of excitons\cite{Kavokin_ExcitonsMediatedSC_NatureMaterials2016,Little_OrganicSCexcitons_PR1964,Little_ExcitonMechanismSC_JPolSci1970,Allender_ExcitonMechanismSC_PRB1973} or plasmons~\cite{Takada_plasmonicSC_JPSJ1978,Rietschel_RoleCoulombSC_PRB1983,Frohlich_SuperconductiviyInnesShellPlasmons_JPC1968}. In particular, the plasmon mechanism has been extensively investigated arguing that it could induce or significantly enhance superconductivity in many and very different classes of systems. These include perovskitic oxides~\cite{Takada_SCinSrTiO3_JPSJ1980,Kresin_TcSClowDim_PRB1987,Kresin_PlasmonCuprates_PhysC1988}, metalchloronitrides~\cite{Bill_DynScreeningMClN_PRB2002,Akashi_HfLiMNCl_PRB2012}, organic superconductors~\cite{Bill_ElectronicCollectiveModes_LayeredSC_PRB2003,Akashi_dopedFullerene_PRB2013} and light-element systems such as lithium metal and high pressure hydride superconductors~
\cite{Akashi_PlasmonSCDFT_PRL2013,Akashi_SH3_PRB2015,Sano_VanHove_SH3_PRB2016,Arita_NonEmpiricalLightElement_AdvMat2017}.

For conventional superconductors, calculations of \tc\ are commonly based on Eliashberg theory~\cite{Migdal1958,Eliashberg1960,VonsovskySuperconductivityTransitionMetals,AllenMitrovic_TheoryofSuperconductingTc_1983}. This is, in principle, a comprehensive theory of the superconducting state, including both electron-phonon and Coulomb effects. The usual application of Eliashberg theory to realistic systems is, however, oversimplified~\cite{ScalapinoSchriefferWilkins} in that the Coulomb interaction is assumed not to favor Cooper-pair formation, and is reduced to a single parameter $\mu^{*}$~\cite{Morel1962,ScalapinoSchriefferWilkins,AllenMitrovic_TheoryofSuperconductingTc_1983}. The standard Eliashberg framework, thus, is not suitable for a quantitative description of superconductivity supported by electronic mechanisms. Unlike Eliashberg theory, the extension of density functional theory to superconductors~\cite{OGK_SCDFT_PRL1988} (SCDFT) does not involve any semi-empirical approximation for the Coulomb interaction, and enables calculations of \tc\ entirely from first-principles. Nevertheless, SCDFT was formulated to address conventional superconductivity~\cite{LuedersSCDFTI2005,MarquesSCDFTII2005}, so that it employs a static screening of the Coulomb repulsion~\cite{Massidda_Sanna_SUST}. Recently, a generalization of SCDFT for applications to plasmonic superconductivity has been proposed~\cite{Akashi_PlasmonSCDFT_PRL2013,Akashi_SCDFTplasmons_JPSJ2014}. However, in this theory plasmonic effects are included in the superconducting state, but neglected in the normal state.

In this work we extend Eliashberg theory (Secs.~\ref{sec:EliashbergTheory}) and SCDFT (Sec.~\ref{sec:SCDFT}) to provide \emph{ab initio} calculations of plasmonic effects on the superconducting properties of real materials. 
In both frameworks retardation effects in the phonon-mediated and screened Coulomb interactions are treated on the same footing by keeping their characteristic frequency dependence. By applying these methods in Sec.~\ref{sec:Results}, we study how the plasmon contribution affects phonon-induced superconductivity for a set of materials representing the main families of conventional superconductors.

\section{\label{sec:EliashbergTheory} Eliashberg Theory}

Eliashberg theory is a many-body perturbative approach for the description of conventional superconductors, where the pairing is driven by a phonon-induced attraction~\cite{AllenMitrovic_TheoryofSuperconductingTc_1983,VonsovskySuperconductivityTransitionMetals,Ummarino_Eliashberg_2013,Sanna_PolaronsCaC6_PRB2012}.
The method employs the Nambu-Gor'kov's formalism, i.e., the diagrammatic expansion is formulated in terms of an extended, $2 \times 2$, electron Green's function $\bar{G}$ with normal (diagonal, $G$) and anomalous (off-diagonal, $F$) components describing, respectively, single-particle electronic excitations and Cooper pairs. The matrix Green's function is determined via the Dyson's equation:
\begin{equation}
\label{eq:Dyson}
\bar{G}^{-1}(k,i\omega_n)= \bar{G}_0^{-1}(k,i\omega_n)-\bar{\Sigma}(k,i\omega_n),
\end{equation}
where $\bar{G}_0$ is the normal-state Green's function of the noninteracting electron system and $\bar{\Sigma}(k,i\omega_n)=\bar{\Sigma}^{c}(k,i\omega_n)+\bar{\Sigma}^{ph}(k,i\omega_n)$ is the electron self-energy associated with the screened Coulomb and phonon-mediated interactions.
$\bar{G}_0$ can be constructed from the Kohn-Sham (KS) states $\ket{k} \equiv \ket{\boldsymbol{k}l}$ and eigenvalues $\varepsilon_k$ of density functional theory (DFT) in the usual form
\begin{equation}
\bar{G}_0(k,i\omega_n)= \left[i\omega_n \tau_0-\varepsilon_k \tau_3\right]^{-1},
\end{equation}
where $\tau_{\{0,\dots,3\}}$ are the Pauli matrices and the energy $\varepsilon_k$ is measured relative to the chemical potential.

The pairing mechanism is dominated by the phonon-mediated interaction that, being retarded, overcomes the (almost instantaneous) Coulomb repulsion between the electrons. Since the phonon energy scale, set by the Debye frequency $\omega_D$, is much smaller than the electronic Fermi energy $E_F$, the method relies on Migdal's theorem to treat the electron-phonon interaction accurately to order $\omega_D/E_F$. The key approximation consists in retaining for $\bar{\Sigma}^{ph}(k,i\omega_n)$
only the diagram for the dressed phonon exchange by the self-consistently dressed electron propagator ($\bar{G}W$). Due to the absence of an analogous theorem, the treatment of the Coulomb interaction is much harder. On the other hand, the possibility of a Coulomb enhancement of the \tc\ is neglected. 
Coulomb effects are largely accounted for by normal state parameters, i.e., the electron and phonon quasiparticle energies, $\varepsilon_k$ and $\omega_{\boldsymbol{q}\nu}$, and the screened electron-phonon coupling $g_{kk'\nu}$. In addition, there remains a static screened Coulomb repulsion $W(k,k')$, which counteracts superconductivity.
Within the Eliashberg approximation, the phonon and Coulomb contributions to the electron self-energy read, respectively, as
\begin{equation}
\label{eq:Sigmaph}
\bar{\Sigma}^{ph}(k,i\omega_n)=\frac{T}{N(0)}\sum_{k'n'}\tau_3 \bar{G}(k',i\omega_{n'})\tau_3\, \lambda_{k,k'}(i\omega_n-i\omega_{n'})
\end{equation}
and
\begin{equation}
\label{eq:Sigmac}
\bar{\Sigma}^{c}(k,i\omega_n)=-T\sum_{k'n'}\tau_3 \bar{G}(k',i\omega_{n'})\tau_3 W(k,k')-v^{xc}_k \tau_3.
\end{equation}
Following the standard practice~\cite{AllenMitrovic_TheoryofSuperconductingTc_1983}, the anisotropic electron-phonon coupling $\lambda_{k,k'}(i\nu_n)$ in Eq.~(\ref{eq:Sigmaph}) is defined by the spectral representation
\begin{equation}
\lambda_{k,k'}(i\nu_n)=\int_0^{\infty} d\omega\,\alpha^2F_{k,k'}(\omega) \frac{2\omega}{\nu^2_n+\omega^2},\label{eq:Spectralreprlambda}
\end{equation}
in terms of the Eliashberg function
\begin{equation}
\alpha^2F_{k,k'}(\omega)=N(0)\sum_{\nu} |g_{kk'\nu}|^2\delta(\omega-\omega_{\boldsymbol{q}\nu}),
\end{equation}
where $N(0)$ is the electronic density of states at the Fermi level.
Eq.~(\ref{eq:Sigmac}) includes the subtraction of the exchange-correlation potential of KS-DFT, $v^{xc}$, so that the resulting Coulomb self-energy is purely off-diagonal. This prevents one from double counting Coulomb effects in the normal state, which are already included in the KS band structure $\varepsilon_k$ entering $\bar{G}_0$.

The total self-energy is more conveniently rewritten in terms of three scalar functions given by the coefficients of the Pauli matrix representation for $\bar{\Sigma}$:  
\begin{align}
\bar{\Sigma}(k,i\omega_n)=&i\omega_n\left[1-Z(k,i\omega_n)\right]\tau_0+\left[\chi(k,i\omega_n)-v^{xc}_k\right]\tau_3\nonumber\\
&+ \phi(k,i\omega_n)\tau_1.\label{eq:Sigmamatrix} 
\end{align}
These are the mass renormalization function $Z(k,i\omega_n)$, the energy shift $\chi(k,i\omega_n)$, and the order parameter $\phi(k,i\omega_n)$. Through the Dyson's equation~(\ref{eq:Dyson}), the calculation of $\bar{G}$ is reduced to solving three coupled equations for $Z$, $\chi$ and $\phi$. In particular, the function $\Delta(k,i\omega_n)=\phi(k,i\omega_n)/Z(k,i\omega_n)$ plays the role of the superconducting energy gap, whereas the quantity $\chi(k,i\omega_n)$ leads to a shift of the chemical potential, which
little affects the formation of the superconducting state. 
By neglecting $\chi$, the equations of interest for the $\tau_0$ and $\tau_1$ components of the Eliashberg self-energy take the form 
\begin{align}
&Z(k,i\omega_n)=1+T\sum_{k',n'}\frac{\lambda_{k,k'}(i\omega_n-i\omega_{n'})}{N(0)}\frac{\omega_{n'}Z(k',i\omega_{n'})}{\omega_n\Theta(k',i\omega_{n'})},\label{eq:ZEliashberg}\\
&\phi(k,i\omega_n)=T\sum_{k',n'}\left[\frac{\lambda_{k,k'}(i\omega_n-i\omega_{n'})}{N(0)}-W(k,k')\right]\nonumber\\
&\quad\quad\quad\quad\quad\times\frac{\phi(k',i\omega_{n'})}{\Theta(k',i\omega_{n'})}\label{eq:phiEliashberg},
\end{align}
where
\begin{equation}
\Theta(k,i\omega_n)=\left[\omega_n Z(k,i\omega_n)\right]^2+\varepsilon_k^2+\phi^2(k,i\omega_n).  
\end{equation}
Note that, since retardation effects in the Coulomb repulsion are disregarded, $Z(k,i\omega_n)$ is entirely determined by the phonon-mediated interaction~\cite{AllenMitrovic_TheoryofSuperconductingTc_1983}. Moreover, the Coulomb contribution to $\phi(k,i\omega_n)$, given by the second term of Eq.~(\ref{eq:phiEliashberg}), is frequency independent. 

Several approximations are commonly employed in order to reduce the workload involved in solving the Eliashberg equations~(\ref{eq:ZEliashberg}) and~(\ref{eq:phiEliashberg}). Essentially, since the superconducting pairing occurs mainly within an energy window $\sim \omega_D$ around the Fermi surface, the equations are simplified by averaging over $k$ and $k'$ in the electronic states on the Fermi surface as follows:
\begin{equation}
\label{eq:Fermiaverage}
f \equiv \langle \langle f(k,k')\rangle \rangle_{FS}=\frac{1}{N(0)^2} \sum_{k,k'} f(k,k')\delta(\varepsilon_{k})\delta(\varepsilon_{k'}).
\end{equation}
Although being quite accurate for phonons, this approximation is not justified for the Coulomb interaction, which may remain large for energies up to $E_F$. In practice, following the arguments of Morel and Anderson~\cite{MorelAnderson}, $W(k,k')$ in Eq.~(\ref{eq:phiEliashberg}) can be replaced by a strongly-reduced pseudo-potential $\mu*$, with an energy cutoff $\omega_c\sim 10 \omega_D$, which effectively accounts for the Coulomb scattering of electrons far from the Fermi surface.
The Morel-Anderson pseudo-potential is defined by the expression
\begin{equation}
\mu^*=\frac{\mu}{1+\mu \log{(E_F/\omega_c)}}, \end{equation}
where $\mu \equiv N(0)\langle \langle W(k,k')\rangle \rangle_{FS}$. 
In most applications, however, $\mu^*$ is treated as a semi-empirical parameter fitted as to reproduce the experimental critical temperature. With the above mentioned approximations, the Eliashberg approach involves solving numerically the following isotropic equations:
\begin{align}
&\left[1-Z(i\omega_n)\right]i\omega_n=-\frac{T}{N(0)}\sum_{n'}\lambda(i\omega_n-i\omega_{n'})\nonumber\\
&\qquad\qquad\qquad\qquad\times\int d\varepsilon N(\varepsilon)\frac{i\omega_{n'}Z(\varepsilon,i\omega_{n'})}{\Theta(\varepsilon,i\omega_{n'})},\\
&\phi(i\omega_n)=\frac{T}{N(0)}\sum_{n'}\left[\lambda(i\omega_n-i\omega_{n'})-\mu^*\theta(\omega_c-|\omega_{n'}|)\right]\nonumber\\
&\qquad\qquad\times \int d\varepsilon N(\varepsilon)\frac{\phi(\varepsilon,i\omega_{n'})}{\Theta(\varepsilon,i\omega_{n'})}.
\end{align}


\begin{figure}
\includegraphics[width=\columnwidth]{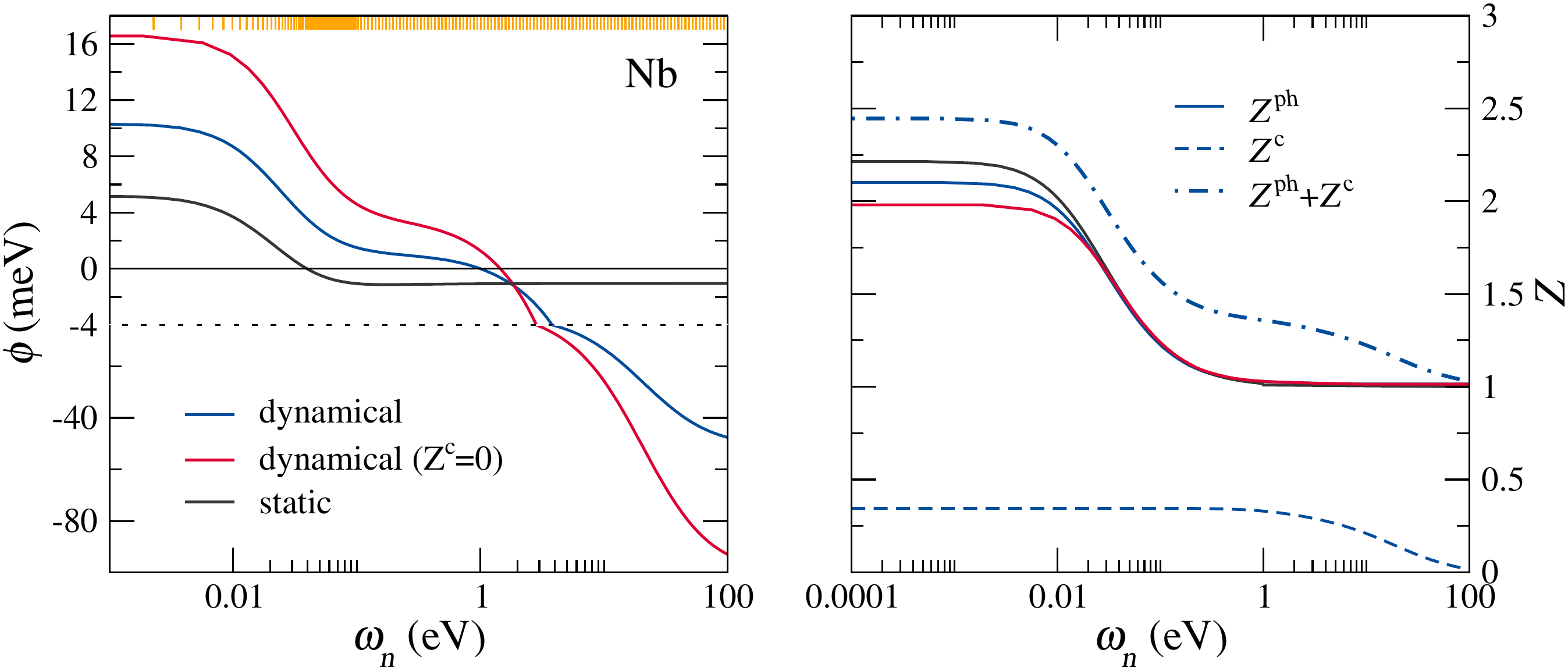} 
\caption{Left: Eliashberg superconducting gap function $\phi$ for bulk Nb in the full dynamical (blue), weak-coupling dynamical (red) and static (black) Coulomb approach. Right: Eliashberg mass renormalization function $Z$ and its decomposition (for the dynamical case) into Coulomb $Z^{c}$ and phononic $Z^{ph}$ components. All the quantities are computed in the low temperature limit.
}\label{fig:phi_and_Z_niobium_example}
\end{figure}

\subsection{Plasmonic extension of Eliashberg Theory}\label{sec:PlasmonicEliashberg}

We go beyond Eq.~(\ref{eq:Sigmac}) for the Coulomb self-energy by assuming the $\bar{G}W$ approximation, i.e., we still neglect vertex corrections, but introduce the dynamical screening of the Coulomb interaction through the frequency-dependent dielectric function. Hence, we consider the self-energy expression
\begin{align}
\label{eq:SigmacGW}
\Sigma^{c}(k,i\omega_n)=&-T\sum_{k'n'}\tau_3 \bar{G}(k',i\omega_{n'})\tau_3 W_{k,k'}(i\omega_n-i\omega_{n'})\nonumber\\
&-v^{xc}_k \tau_3,
\end{align}
where the Coulomb potential $W_{k,k'}(i\nu_n)$ is obtained from the symmetrized dielectric function $\epsilon_{\boldsymbol{G}\boldsymbol{G}'}(\boldsymbol{q},i\nu_n)$ as
\begin{equation}
\label{eq:Wdyn}
W_{k,k'}(i\nu_n)=\frac{4\pi}{\Omega}\sum_{\boldsymbol{G}\boldsymbol{G}'}\frac{\epsilon^{-1}_{\boldsymbol{G}\boldsymbol{G}'}(\boldsymbol{q},i\nu_n)\rho^k_{k'}(\boldsymbol{G})\rho^{k\,*}_{k'}(\boldsymbol{G}')}{|\boldsymbol{q}+\boldsymbol{G}||\boldsymbol{q}+\boldsymbol{G}'| }.              
\end{equation}
Here, $\Omega$ is the unit cell volume, $\boldsymbol{G}$ are reciprocal lattice vectors and $\boldsymbol{q}$ is the difference $\boldsymbol{k}-\boldsymbol{k}'$ reduced to the first Brillouin zone. The symmetric form of the dielectric function is defined by
\begin{equation}
\epsilon^{-1}_{\boldsymbol{G}\boldsymbol{G}'}(\boldsymbol{q},i\nu_n)=\delta_{\boldsymbol{G}\boldsymbol{G}'}+\frac{\sqrt{4\pi}}{|\boldsymbol{q}+\boldsymbol{G}|}\chi_{\boldsymbol{G}\boldsymbol{G}'}(\boldsymbol{q},i\nu_n)\frac{\sqrt{4\pi}}{|\boldsymbol{q}+\boldsymbol{G}'|},  
\end{equation}
where $\chi_{\boldsymbol{G}\boldsymbol{G}'}(\boldsymbol{q},i\nu_n)$ is the reducible polarization. The pair density matrix elements read as $\rho^k_{k'}(\boldsymbol{G})\equiv\braket{k'|e^{-i(\boldsymbol{q}+\boldsymbol{G})\cdot \boldsymbol{r}}|k}$.

For numerical convenience, we rewrite $W_{k,k'}(i\nu_n)$ in the form
\begin{equation}
W_{k,k'}(i\nu_n)=V(k,k')+\int_0^{\infty} d\omega\, \mathcal{S}_{k,k'}(\omega) \frac{2\omega}{\omega^2+\nu_n^2},\label{eq:SpectralreprW}
\end{equation}
where $V(k,k')=\displaystyle{\frac{4\pi}{\Omega}\sum_{\boldsymbol{G}}\frac{|\braket{k'|e^{-i(\boldsymbol{q}+\boldsymbol{G})\cdot \boldsymbol{r}}|k}|^2}{|\boldsymbol{q}+\boldsymbol{G}|^2}}$ are the matrix elements of the bare Coulomb interaction and
\begin{align}
\mathcal{S}_{k,k'}(\omega)=\frac{16\pi}{\Omega}\Im \sum_{\boldsymbol{G}\boldsymbol{G}'} \frac{\chi_{\boldsymbol{G}\boldsymbol{G}'}(\boldsymbol{q},\omega+i\eta)\rho^k_{k'}(\boldsymbol{G})\rho^{k\,*}_{k'}(\boldsymbol{G}')}{|\boldsymbol{q}+\boldsymbol{G}|^2|\boldsymbol{q}+\boldsymbol{G}'|^2 } 
\end{align}
is the spectral function of the electronic polarization.
Note that the second term on the right hand side of Eq.~(\ref{eq:SpectralreprW}) is formally equivalent to the spectral representation of the anisotropic electron-phonon coupling (Eq.~(\ref{eq:Spectralreprlambda})). 
The screened Coulomb potential in its spectral representation can be separated into a static and a dynamical part,
\begin{equation}
\label{eq:decompositionW}
 W_{k,k'}(i\nu_n)=W(k,k')+\Delta W_{k,k'}(i\nu_n),
\end{equation}
where the latter, given by
\begin{equation}
\Delta W_{k,k'}(i\nu_n)=\int_0^{\infty} d\omega\, \mathcal{S}_{k,k'}(\omega) \left[\frac{2\omega}{\omega^2+\nu_n^2}- \frac{2} {\omega}\right],
\end{equation}
incorporates plasma oscillations. We approximate Eq.~(\ref{eq:decompositionW}) by its average taken over the corresponding surfaces of constant energy, $\varepsilon$, in $\boldsymbol{k}$-space as
\begin{align}
&W(\varepsilon,\varepsilon')=\frac{1}{N(\varepsilon)N(\varepsilon')} \sum_{k,k'} W(k,k')\delta(\varepsilon_k-\varepsilon)\delta(\varepsilon_{k'}-\varepsilon'),\label{eq:Waverage}\\
&\mathcal{S}(\varepsilon,\varepsilon',\omega)=\frac{1}{N(\varepsilon)N(\varepsilon')} \sum_{k,k'}
\mathcal{S}_{k,k'}(\omega) \delta(\varepsilon_k-\varepsilon)\delta(\varepsilon_{k'}-\varepsilon').\label{eq:DeltaWaverage}
\end{align}

It should be observed that Eq.~(\ref{eq:Waverage}) is a generalization of Eq.~(\ref{eq:Fermiaverage}) for the conventional isotropic Eliashberg theory. By using Eqs.~(\ref{eq:Waverage}) and~(\ref{eq:DeltaWaverage}) for the Coulomb interaction in the expression for the self-energy, we obtain the following Coulomb contributions to the Eliashberg functions $Z$ and $\phi$:
\begin{align}
&Z^c(\varepsilon,i\omega_n)=-T \sum_{n'}\int d\varepsilon' N(\varepsilon')\int d\omega \mathcal{S}(\varepsilon,\varepsilon',\omega)\nonumber\\
&\quad\quad\quad\quad\quad\times \frac{2\omega}{\omega^2+(\omega_n-\omega_{n'})^2}\,\frac{\omega_{n'}Z(\varepsilon',i\omega_{n'})}{\omega_n\Theta(\varepsilon',i\omega_{n'})},\label{eq:ZcEli}\\
&\phi^c(\varepsilon,i\omega_n)=-T\sum_{n'}\int d\varepsilon' N(\varepsilon')\biggl\{W(\varepsilon,\varepsilon')\nonumber\\
&\,\,+\int d\omega \mathcal{S}(\varepsilon,\varepsilon',\omega)\left[\frac{2\omega}{\omega^2+(\omega_n-\omega_{n'})^2}- \frac{2} {\omega}\right]\biggr\}\frac{\phi(\varepsilon',i\omega_{n'})}{\Theta(\varepsilon',i\omega_{n'})}.\label{eq:phicEli}
\end{align}
The influence of these terms on \tc\ can be easily seen by considering that in the simple BCS limit one has \tc$ \propto \omega_D \exp\left[-(1+\lambda)/(\lambda-\mu*)\right]$, where $1+\lambda$ comes from the electron-phonon $Z$ term in the Eliashberg equations. Here, the dynamical contribution to the anomalous kernel $\phi^c$ (given by the second term of Eq.~(\ref{eq:phicEli})) enhances the Coulomb repulsion $\mu$ between the electrons in the energy scale of the plasmon frequency $\omega_{pl}$. On the other hand, since $\omega_D \ll \omega_{pl}$, the effective Coulomb repulsion decreases from the original value $\mu*$, favoring superconductivity (higher \tc ). This effect, however, is counteracted by the Coulomb correction $Z^c$ to the effective mass, which adds up to the phononic term $(1+\lambda)$, contributing to the reduction of \tc .

Fig.~\ref{fig:phi_and_Z_niobium_example} shows the Matsubara frequency dependence of the mass renormalization $Z$ and gap function $\phi$ for bulk Nb. The inclusion of retardation effects in the Coulomb interaction leads in $\phi$ to large negative tails at high energy. Since the high-energy gap function is negative, the plasmonic coupling serves as an effective attraction, which, according to Eq.~(\ref{eq:phicEli}), increases the value of $\phi$ at the Fermi level. This effect is less pronounced when $Z^c$ is included.
The right panel of Fig.~\ref{fig:phi_and_Z_niobium_example} shows the decomposition of $Z$ into phononic and Coulomb contributions. $Z^{ph}$ has a peak at low frequency with energy width of the order of the Debye energy and above $\omega_D$ converges to 1. $Z^c$, instead, which is non-zero only in the dynamical approach, decays at the plasmonic energy scale.

As evident from Fig.~\ref{fig:phi_and_Z_niobium_example}, the main difficulty in solving the Eliashberg Eqs.~(\ref{eq:ZcEli}) and~(\ref{eq:phicEli}) is related to the fact that the integration (both in $\omega_n$ and $\varepsilon$) has to be performed on a huge energy scale, and therefore cannot be tackled by brute force computation. Just to give an indication, at $T=2.8~K$, the number of Matsubara points within the plotted energy range is of the order of 70 thousand, and reaching a tight convergence would require an even larger energy window of several hundred eV. To overcome this slow convergence problem in the numerical implementation of the equations we have adopted the following strategies: 
i) We have used a logarithmic $\varepsilon$ integration mesh. This allows for a dense discretization at low energy, where variations in the functions have to be accounted for more accurately, but extends up to arbitrarily large energies with relatively few additional points.
ii) Similarly, we have adopted a non-homogeneous mesh of Matsubara points. Since Matsubara frequencies are fixed by the temperature, a non-linear mesh can be obtained by pruning points and redistributing their weight. The resulting Matsubara mesh at 2.8~K is indicated by orange ticks in the left panel of Fig.~\ref{fig:phi_and_Z_niobium_example}.
iii) The dynamical Coulomb interaction itself depends on the (bosonic) Matsubara frequencies and on the energy. When computing the interaction from first principles, a huge computational cost is associated with the calculation of the matrix elements of the dielectric function at high frequency, with respect to the KS states at high energy. For this reason, we have introduced high-energy cutoffs in $\omega_n$ and $\varepsilon$ (typically of the order of 50-100~eV). Above this energy, the dielectric function of the material is replaced with that of the homogeneous electron gas in the plasmon-pole approximation. The parameters which enter the latter are fitted to the explicitly computed values of $\mathcal{S}(\varepsilon,\varepsilon',\omega)$ at the cutoff, so to ensure a good overall match to the actual interaction. This approach not only reduces the numerical cost in computing the interaction, but also allows for the analytical integration of the Matsubara frequencies from the cutoff energy to infinity. We point out that these techniques do not introduce additional errors in the method. However, they involve convergence parameters that have to be carefully chosen in order to achieve the correct numerical result.


\section{\label{sec:SCDFT} Density Functional Theory for Superconductors}
Density functional theory for superconductors (SCDFT) is an extension of conventional DFT for \emph{ab initio} calculations of material-specific
properties in the superconducting state~\cite{OGK_SCDFT_PRL1988}.
The theory includes the superconducting order parameter $\chi_{sc}(\mathbf{r},\mathbf{r}')$ as an additional density. The corresponding non-interacting KS system then reproduces, in principle exactly, both the normal density and the superconducting order parameter of the real system. In the so-called decoupling approximation (on which Eliashberg theory is also based), the KS system is fully determined by solving the BCS-like gap equation
\begin{equation}
\label{eq:SCDFTgapeq}
\Delta_{s\,k}=-\mathcal{Z}_{k}\Delta_{s\,k}-\frac{1}{2}\sum_{k'}\mathcal{K}_{k,k'}\frac{\tanh\left(\frac{\beta}{2}E_{k'}\right)}{E_{k'}}\Delta_{s\,k'},
\end{equation}
where $E_{k}=\sqrt{\varepsilon^2_{k}+|\Delta_{s\,k}|^2}$ are the KS excitation energies and $\beta$ is the inverse temperature.
The kernel of the equation consists of a diagonal part, $\mathcal{Z}_k=\mathcal{Z}^{ph}_k$, and a nondiagonal part, $\mathcal{K}_{k,k'}$. $\mathcal{Z}_k^{ph}$ plays the role of the renormalization function in the Eliashberg equations, whereas $\mathcal{K}_{k,k'}=\mathcal{K}_{k,k'}^{c}+\mathcal{K}_{k,k'}^{ph}$, which includes both Coulomb and phonon-mediated effects, is responsible for the binding of the electrons in Cooper pairs. Compared to Eliashberg theory, SCDFT features two major advantages: (i) the treatment of the Coulomb repulsion does not resort on any empirical parameter $\mu*$, (ii) all the Matsubara frequency summations are evaluated analytically in the construction of the exchange-correlation (xc) kernels. As in Eliashberg theory, phonon dynamics is properly included, but at the same time the gap equation retains the form of a static BCS equation. Hence, Eq.~(\ref{eq:SCDFTgapeq}) allows one to account for the full anisotropy of materials at a low computational cost. However, the accuracy of the method is bound by the quality of the available functionals.

Making a connection to many-body perturbation theory, approximate xc kernels have been derived from approximations for the xc self-energy operator, via the Sham-Schl\"uter equation in Nambu space. The first SCDFT functional by L\"uders, Marques and co-workers (LM)~\cite{LuedersSCDFTI2005,MarquesSCDFTII2005} employed the $\bar{G}_sW$ approximation for the self-energy in the statically screened Coulomb repulsion and phonon-mediated interaction. By construction, this functional neglected higher order processes included in Eliashberg theory by the self-consistent dressing of the electron Green's function in the $\bar{G}W$ self-energy. The LM approximation, thus, was not validated by Migdal's theorem, which made it of questionable accuracy for treating electron-phonon coupling effects. To solve this issue, Sanna, Pellegrini and Gross (SPG)~\cite{SPG_EliashbergSCDFT_PRL2020} have recently introduced a parametrization of the functional based on the electron-phonon Eliashberg self-energy for a simplified (Einstein) phonon spectrum. The new SPG kernels, give superconducting transition temperatures and gaps in excellent agreement with experiments~\cite{Livas_PH_PRB2016,Livas_SH3_EPJB2016,Livas_SH3_EPJB2016,Livas_PunderPress_PRM2017,Monni_SunderPress_PRB2017,Pellegrini_WO3_PRM2019,Errea_LaH10Nature2020}, while still having a simple analytic form. 

Further extensions and applications of the LM functional have addressed the description of superconductivity in the presence of magnetic fields~\cite{Linscheid_SpinSCDFT_I_PRB2015,Linscheid_SpinSCDFT_II_PRB2015} and in real space~\cite{Linscheid_OP_PRL2015}, the inclusion of spin-fluctuations contributions to the pairing~\cite{EssenbergerPRB2014,Essenberger_FeSe_PRB2016} and the treatment of the dynamical screening of the Coulomb interaction~\cite{Akashi_PlasmonSCDFT_PRL2013,Akashi_SCDFTplasmons_JPSJ2014,Akashi_dopedFullerene_PRB2013,Akashi_HfLiMNCl_PRB2012,Akashi_SH3_PRB2015}. In a first attempt to introduce plasmonic effects in SCDFT, Akashi and Arita~\cite{Akashi_SCDFTplasmons_JPSJ2014,Akashi_PlasmonSCDFT_PRL2013} have proposed a dynamical correction to the pairing kernel $\mathcal{K}_{k,k'}^{c}$ by retaining the frequency dependence of the Coulomb interaction at the RPA level in the exchange anomalous self-energy. The method, implemented in the multipole plasmon approximation, has given a systematic increase of the \tc\ by $10-20\%$ in compressed sulfur hydrates H$_2$S and H$_3$S, and by a factor of $2$ in Al and Li under pressure.

\begin{figure}[h]
\includegraphics[width=\columnwidth]{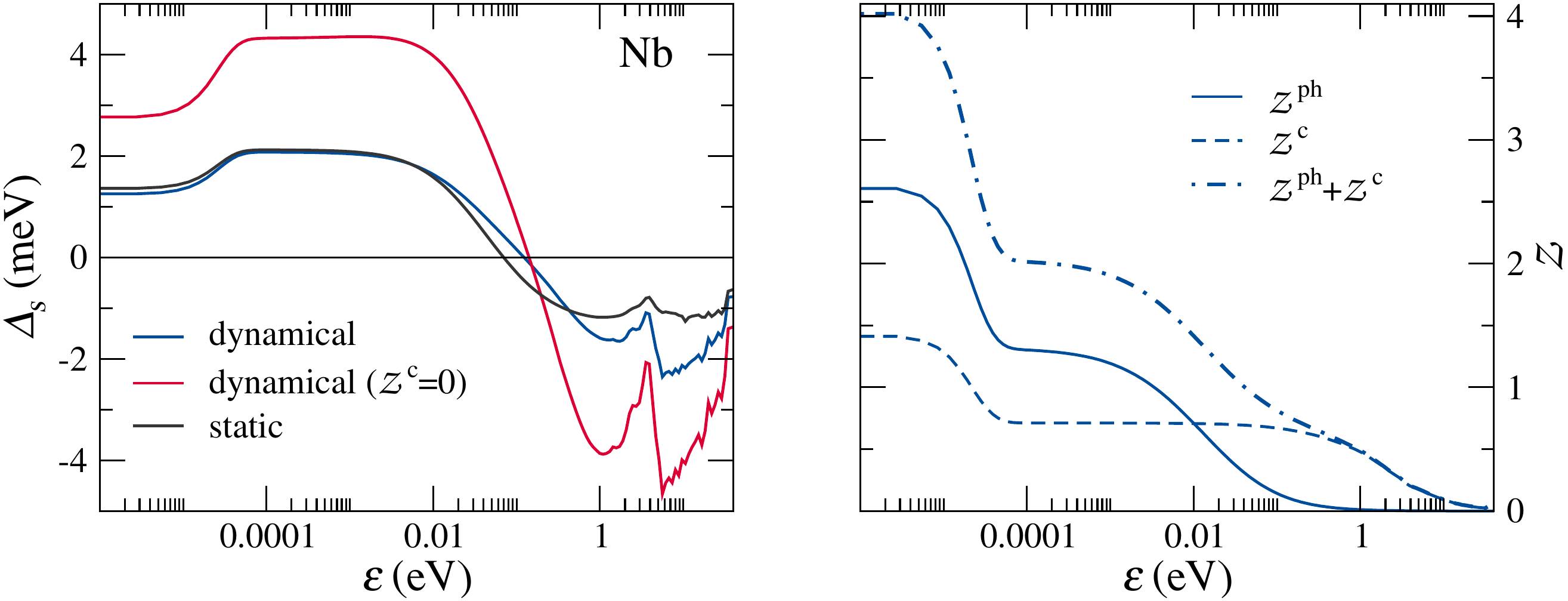}
\caption{Left: SCDFT gap function $\Delta_s$ for bulk Nb in the dynamical (blue), weak-coupling dynamical (red) and static (black) Coulomb approach. Right: SCDFT $\mathcal{Z}$ kernel and its decomposition (for the dynamical case) into Coulomb $\mathcal{Z}^{c}$ and phononic $\mathcal{Z}^{ph}$ components. All the quantities are computed in the low temperature limit and using the SPG phononic functional.
}\label{fig:Ds_and_Zker_niobium_example}
\end{figure}

\subsection{Plasmonic SCDFT with mass term\label{sec:SCDFTplasmonsZTheory}}

The results of plasmonic Eliashberg theory for Nb (Fig.~\ref{fig:phi_and_Z_niobium_example}) suggest that the $Z^c$ term, stemming from the diagonal part of the dynamical Coulomb self-energy, should play a major role in determining \tc . $Z^c$ can be viewed as the Coulomb counterpart of the mass renormalization enhancement, $1+\lambda$, that corrects the BCS predictions to their strong coupling values in Eliashberg theory~\cite{CarbotteRMP,AllenMitrovic_TheoryofSuperconductingTc_1983}. Accordingly, it is expected to be relevant for strong electron-plasmon interactions. As discussed above, the recently developed SCDFT scheme for plasmonic superconductivity neglects this contribution by assuming a null diagonal kernel $\mathcal{Z}^c$ and can, thus, be regarded as a weak-coupling plasmonic theory. In this section we propose a more general SCDFT approach, which also includes plasmonic corrections to the mass enhancement.

By following a procedure analogous to that presented in Ref.~\onlinecite{PhDMarques} for the treatment of the electron-phonon coupling, we construct the SCDFT plasmonic kernels from the $\bar{G}_sW$ self-energy in the screened Coulomb potential. In the isotropic approximation (Eqs.~(\ref{eq:Waverage}) and ~(\ref{eq:DeltaWaverage})), we obtain the following expressions:
\begin{align}
\mathcal{Z}^c\left(\varepsilon\right)=&\frac{1}{\tanh\left(\frac{\beta}{2}\varepsilon\right)}\int d\varepsilon' N\left(\varepsilon'\right)\int d\omega \mathcal{S}\left(\varepsilon,\varepsilon',\omega\right)\nonumber\\ 
&\times \frac{\partial}{\partial\varepsilon} \left[I\left(\varepsilon,\varepsilon',\omega\right)+I\left(\varepsilon,-\varepsilon',\omega\right)\right],\label{eq:ZcSCDFT}\\
\mathcal{K}^c\left(\varepsilon,\varepsilon'\right)=& W\left(\varepsilon,\varepsilon'\right)-2\int d\omega \mathcal{S}\left(\varepsilon,\varepsilon',\omega\right)\nonumber\\
&\times \left[\frac{I(\varepsilon,\varepsilon',\omega)-I(\varepsilon,-\varepsilon',\omega)}{\tanh\left(\frac{\beta}{2}\varepsilon\right)\tanh\left(\frac{\beta}{2}\varepsilon'\right)}+\frac{1}{\omega}\right],\label{eq:KcSCDFT}
\end{align}
where the quantity $I(\varepsilon,\varepsilon',\omega)$ is defined in terms of the Fermi and Bose distribution functions $f$ and $b$ by
\begin{align}
I(\varepsilon,\varepsilon',\omega)=&J(\varepsilon,\varepsilon',\omega)-J(\varepsilon,\varepsilon',-\omega),\\
J(\varepsilon,\varepsilon',\omega)=&\left[f(\varepsilon)+b(\omega)\right]\frac{f(\varepsilon')+ f(\varepsilon-\omega)}{\varepsilon-\varepsilon'-\omega}.
\end{align}

In Fig.~\ref{fig:Ds_and_Zker_niobium_example} we show the energy dependence of the calculated KS gap function and kernel $\mathcal{Z}$ for bulk Nb. As in Eliashberg theory, plasmonic contributions enhance the high-energy negative gap. The effect is much more pronounced in the weak-coupling approach, within which the value of the KS gap at the Fermi level almost doubles compared to the static and full dynamical cases.
The kernels $\mathcal{Z}^c$ and $\mathcal{Z}^{ph}$ on the right panel of Fig.~\ref{fig:Ds_and_Zker_niobium_example} have the shape of two over-imposed peaks. The sharper peak is strongly temperature dependent, and occurs very close to the Fermi level, at $\left| \varepsilon \right| < 10^{-4}eV$. This energy scale is not directly related to that of the couplings, but arises from the constraint that the KS system should reproduce the interacting normal and anomalous densities~\cite{SPG_EliashbergSCDFT_PRL2020}. On the other hand, the broader peak decays on the energy scale of the interactions, i.e., at the plasmon energy in $\mathcal{Z}^c$ and at the Debye energy in $\mathcal{Z}^{ph}$.


\section{Hybrid Eliashberg}\label{sec:hybridEliashbergTheory}

By virtue of Migdal's theorem~\cite{Migdal1958,AllenMitrovic_TheoryofSuperconductingTc_1983}, the $FW$ approximation for the anomalous self-energy in Eliashberg theory describes the phonon-mediated pairing very accurately. 
On the other hand, there is no a priori indication that the $FW$ scheme improves over $F_sW$ for the treatment of plasmonic effects. Here, we consider a hybrid Eliashberg-SCDFT theory in which the Coulomb part of the pairing self-energy is in the $F_sW$ form, where $F_s$ is the KS Green's function which reproduces the superconducting order parameter in the Eliashberg approximation. 

Since the KS system has the same anomalous density of the interacting system, the following equality holds~\cite{SPG_EliashbergSCDFT_PRL2020,LuedersSCDFTI2005,MarquesSCDFTII2005}:
\begin{equation}
\label{eq:chi}
\chi_{sc}\left(\varepsilon\right)=-\frac{1}{\beta}\sum_n F\left(\varepsilon,i\omega_n\right)=-\frac{1}{\beta}
\sum_n F_{s}\left(\varepsilon,i\omega_n\right),
\end{equation}
where $F_{s}\left(\varepsilon,i\omega_n\right)=\displaystyle{\frac{\Delta_s\left(\varepsilon\right)}{\omega^2_n+E^2}}$ with $E=\sqrt{\Delta_s^2+\varepsilon^2}$. Using the Eliashberg Green's function $F\left(\varepsilon,i\omega_n\right)=\phi\left(\varepsilon,i\omega_n\right)/\Theta\left(\varepsilon,i\omega_n\right)$ to compute $\chi_{sc}$ and evaluating the Matsubara frequency summation on the right hand side of Eq.~(\ref{eq:chi}), yields:
\begin{equation}
\label{eq:chiDeltas}
\chi_{sc}\left(\varepsilon\right)=-\frac{\Delta_s\left(\varepsilon\right)}{2\sqrt{\varepsilon^2+\Delta^2_s\left(\varepsilon\right)}}
\tanh{\frac{\beta\sqrt{\varepsilon^2+\Delta^2_s\left(\varepsilon\right)}}{2}},
\end{equation}
which relates the Eliashberg anomalous density to the KS potential. Solving numerically Eq.~(\ref{eq:chiDeltas}) for $\Delta_s$, allows one to uniquely construct the corresponding $F_s$ in the Eliashberg approximation.
$F_s$ is then used as an input for the Eliashberg equation which determines the Coulomb gap function, i.e., Eq.~(\ref{eq:phicEli}) is replaced with
\begin{align}
&\phi^c(\varepsilon,i\omega_n)=-T\sum_{n'}\int d\varepsilon' N(\varepsilon')\biggl\{W(\varepsilon,\varepsilon')\nonumber\\
&\,\,+\int d\omega \mathcal{S}(\varepsilon,\varepsilon',\omega)\left[\frac{2\omega}{\omega^2+(\omega_n-\omega_{n'})^2}- \frac{2} {\omega}\right]\biggr\}\frac{\Delta_s\left(\varepsilon'\right)}{\omega^2_{n'}+E'^2}.
\end{align}

The obtained $F_s$ is then used as an input for the Eliashberg equation which determines the Coulomb gap function, i.e., Eq.~(\ref{eq:phicEli}) is replaced with
\begin{align}
&\phi^c(\varepsilon,i\omega_n)=-T\sum_{n'}\int d\varepsilon' N(\varepsilon')\biggl\{W(\varepsilon,\varepsilon')\nonumber\\
&\,\,+\int d\omega \mathcal{S}(\varepsilon,\varepsilon',\omega)\left[\frac{2\omega}{\omega^2+(\omega_n-\omega_{n'})^2}- \frac{2} {\omega}\right]\biggr\}\frac{\Delta_s\left(\varepsilon'\right)}{\omega^2_{n'}+E'^2}.
\end{align}

\begin{figure}[h]
\begin{center}
\includegraphics[width=1.0\columnwidth]{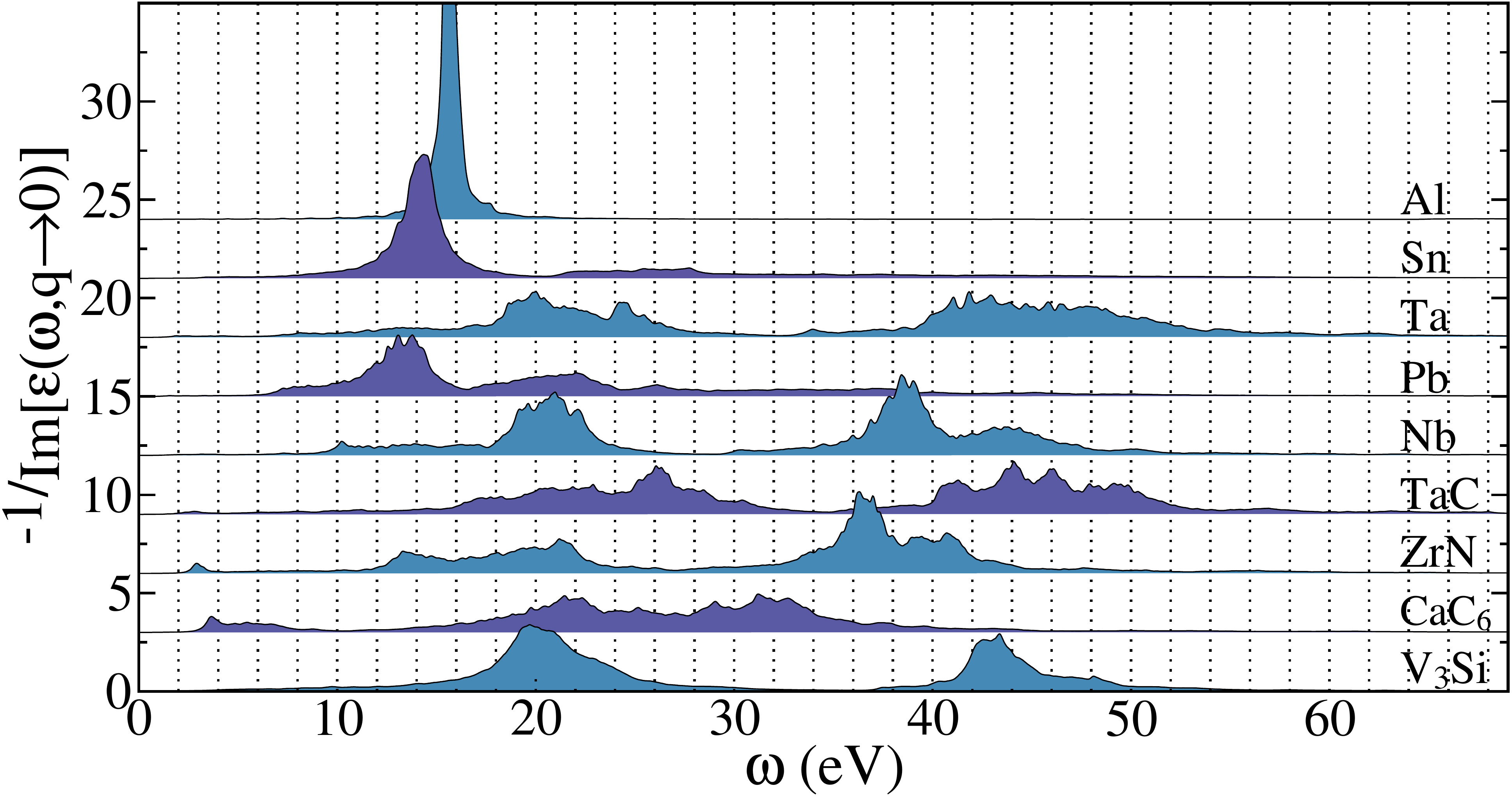}
\caption{Electron energy loss spectra in the low-${\bf q}$ limit, showing the position of the main plasmonic peaks for the chosen set of materials. 
\label{fig:epsm1}}
\end{center}
\end{figure}

\begin{table}[h]
\begin{tabular}{l|rrrrrrrrr}
         & Al & Sn & Ta & Pb & Nb & ZrN & TaC &\x CaC$_{\rm 6}$ & V$_{\rm 3}$Si \\\hline
$\lambda$&\x0.43&\x1.02&\x0.80&\x1.32&\x1.33&\x0.74&\x0.65&\x0.77&\x1.43\\
$Z_0^c$  &0.31&0.28&0.33&0.29&0.33&0.31&0.26&0.35&0.38\\
\tcexp   &1.18&3.8 &4.5 & 7.2& 9.2&
{\large$8.1/ \atop 9.45$} &  {\large$9.7/ \atop 10.2$} &
11.5 &17.0\\
\end{tabular}

\caption{Electron-phonon ($\lambda$) and electron-plasmon ($Z_0^c$) coupling strengths for the test set of materials with associated experimental critical temperatures \tcexp .}\label{tab:Materials}
\end{table}


\section{Computational Results}\label{sec:Results}
\subsection{Material set}\label{sec:Materials}

In order to assess the accuracy of the methods discussed above, we have investigated the effect of the electron-plasmon coupling on the transition temperatures of a set of conventional superconductors. Our set includes experimentally well-characterized systems chosen to cover a wide range of properties, i.e., elemental (Al, Sn, Ta, Nb and Pb) and binary phonon-mediated superconductors (TaC, ZrN, V$_3$Si and CaC$_{\rm 6}$), ranging from weak to strong coupling. To keep the entire procedure \textit{ab initio}, all the calculations have been performed at the theoretical lattice parameters obtained by means of the PBE functional~\cite{PBE}. The dielectric function entering the dynamical Coulomb kernel has been calculated within the random phase approximation (RPA), using the full-potential LAPW code Elk~\cite{elk}. 
Fig.~\ref{fig:epsm1} shows the electron energy loss spectra in the low-${\bf q}$ limit for the chosen set of materials. For both Al and Sn, which are free electron-like metals, one observes, similarly to the homogeneous electron gas, a single pronounced plasmon peak, centered respectively at 16 and 14~eV. All the other systems show a more complex spectrum, varying from the two-peak structure of V$_{\rm 3}$Si, to the broad distributed structures of TaC. Apart from the low-energy plasmons of CaC$_{\rm 6}$ and ZrN, the main plasmonic structures are located at energies above 10~eV.

The electron-phonon and electron-plasmon coupling strengths for the chosen materials are summarized in Tab.~\ref{tab:Materials}, together with the experimental \tc's. The electron-phonon coupling is expressed in terms of the BCS-like coupling constant $\lambda$ defined as the static limit of $\lambda\left(i\nu_n\right)$ in Sec.~\ref{sec:EliashbergTheory}. The electron-plasmon coupling, being strongly energy dependent, cannot be reduced to a simple isotropic parameter, and is thus represented by the energy integrated quantity $Z_0^c$, defined as $Z^c\left(\varepsilon,i\omega_n\right)$ of Eq.~(\ref{eq:ZcEli}) computed at the Fermi level ($\varepsilon=0$) and $\omega_n=0$.

\begin{figure}
\includegraphics[width=\columnwidth]{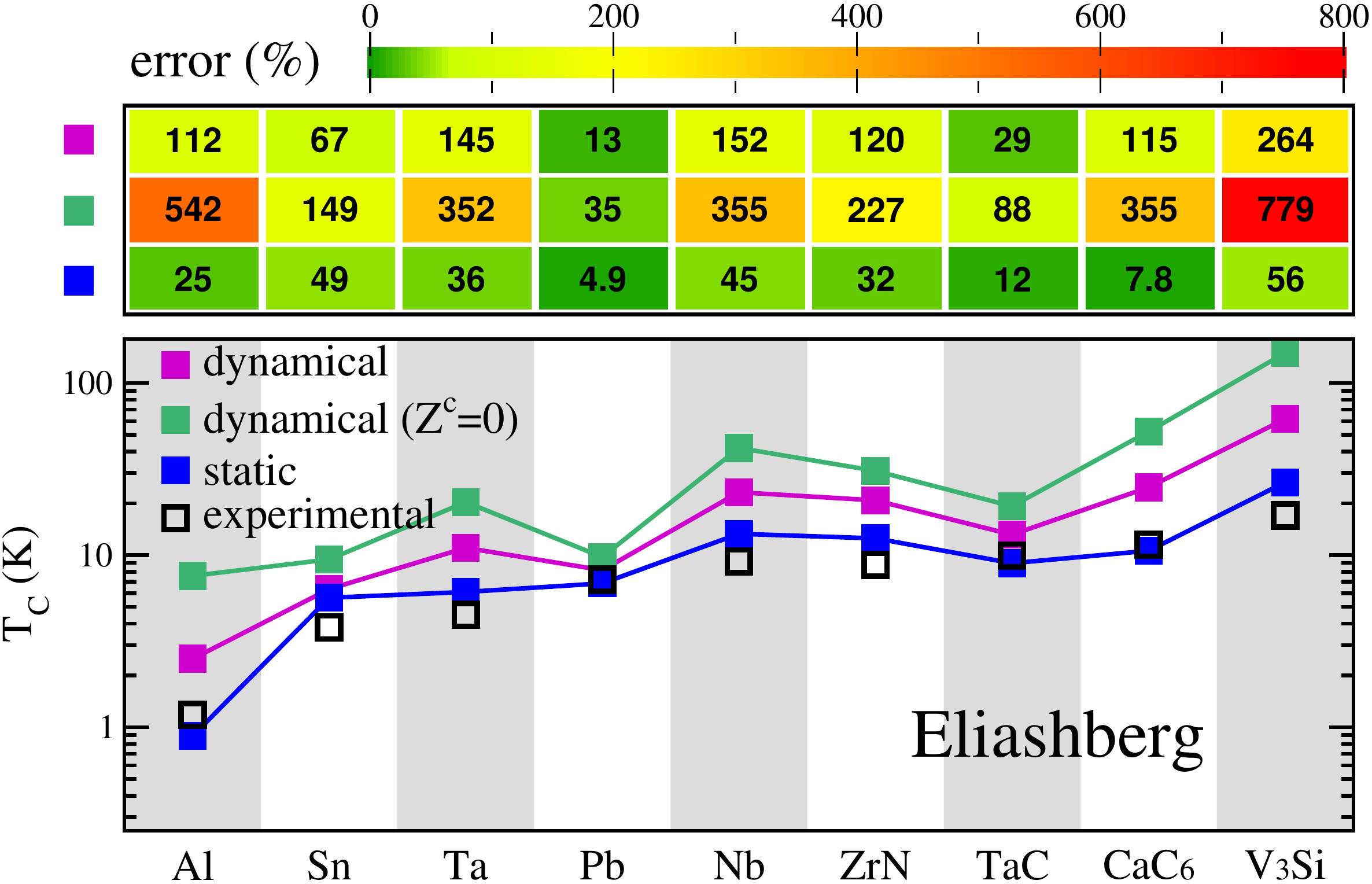}
\caption{Bottom: Eliashberg critical temperatures computed within the full dynamical (violet), weak-coupling dynamical (green) and static (blue) approach for the screened Coulomb interaction. Note the use of a logarithmic scale on the ordinate axis. Top: percentage error (computed with respect to the experimental critical temperature).
Errors are highlighted by a color scale from green (accurate) to yellow and red (large deviation).
}\label{fig:tc_comparisonEli}
\end{figure}


\subsection{Eliashberg}\label{sec:EliashbergResults}

In Fig.~\ref{fig:tc_comparisonEli} and Tab.~\ref{tab:Tc} (columns A-C, L) the experimental values of \tc\ for the chosen set of materials are compared to the values calculated within the Eliashberg approach from Eqs.~(\ref{eq:ZcEli}) and~(\ref{eq:phicEli}) by employing the static and dynamical screening of the Coulomb interaction. For the dynamical case, the weak-coupling results obtained by neglecting the plasmon-induced mass renormalization term $Z^c$ are also shown.

It is evident that the static approximation gives better values for \tc , whereas the plasmonic theory systematically overestimates the experimental data. The inclusion of Coulomb retardation effects in the $\bar{G}W$ approximation yields predicted temperatures that are on average two times bigger than the corresponding experimental values. Notably, the discrepancy between theory and experiments becomes huge when the term $Z^{c}$ is neglected.
Since the inclusion of dynamical screening effects in the Coulomb interaction brings the theory a step closer to being exact, one would expect an improvement in the calculated values of the critical temperature. The apparent worsening of the results can be traced back to the neglect of vertex corrections in the Coulomb self-energy diagram~\cite{Buche_SC_elgas_1990,Takada_vertex_JPSJ1992} and/or the breakdown of the RPA for W. Regarding this point, we should mention that going beyond RPA by using linear-response time-dependent DFT~\cite{PhysRevLett.52.997,Marques2012} within the adiabatic local density approximation in the calculation of the dielectric function does not improve significantly the quality of the results.
These aspects will require further investigations. 
As a matter of fact, our \emph{ab initio} treatment of the static Coulomb interaction in Eliashberg theory appears to be very accurate, confirming previous results along these lines~\cite{Sano_VanHove_SH3_PRB2016,Sanna_GenuinePreictionsEliashberg_JPSJ_2018,Arita_EfficientMEcalculations_Arxiv2020}.


\begin{figure*}
\includegraphics[width=0.485\textwidth]{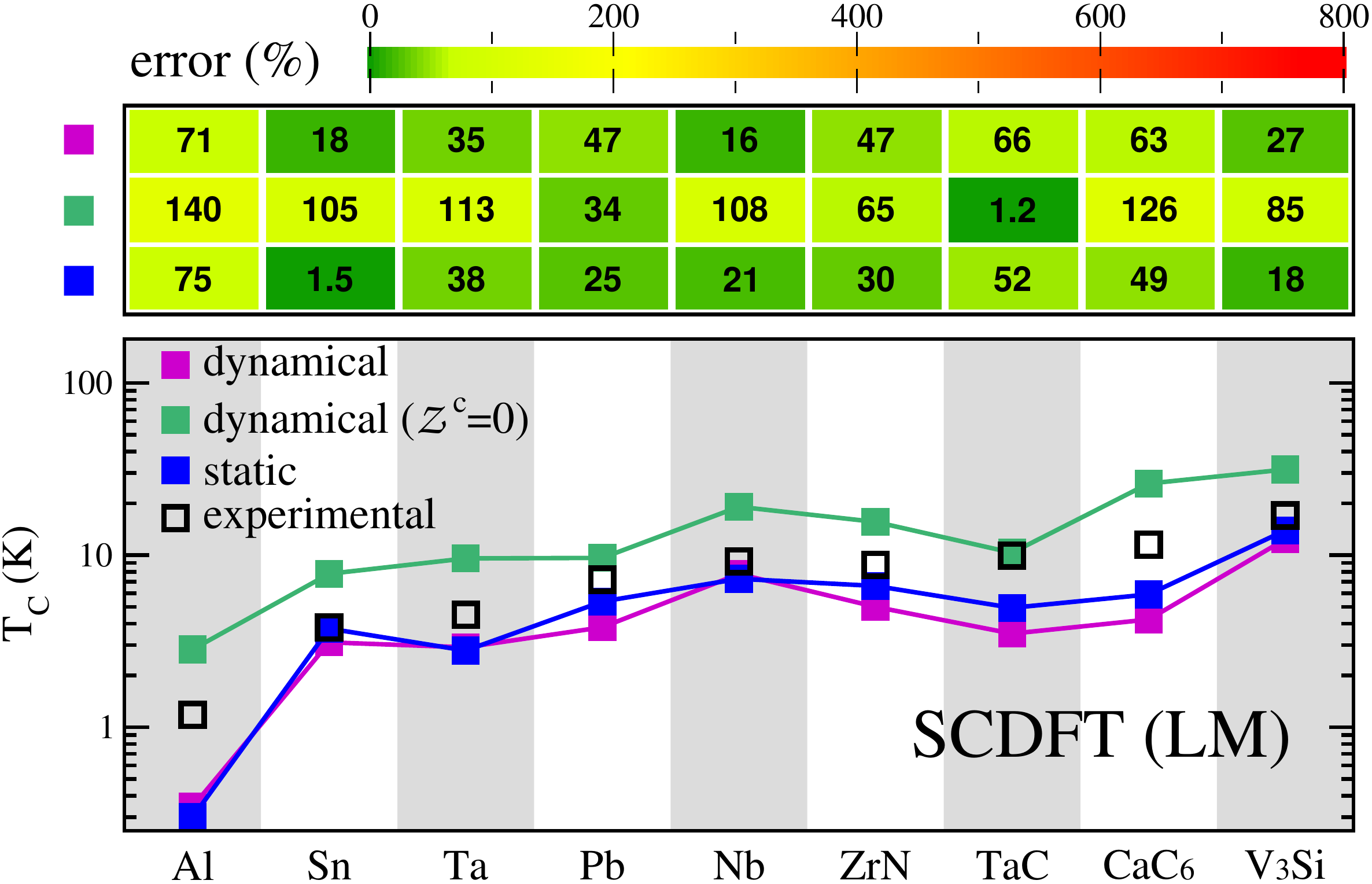}
\hspace{0.01\textwidth}
\includegraphics[width=0.485\textwidth]{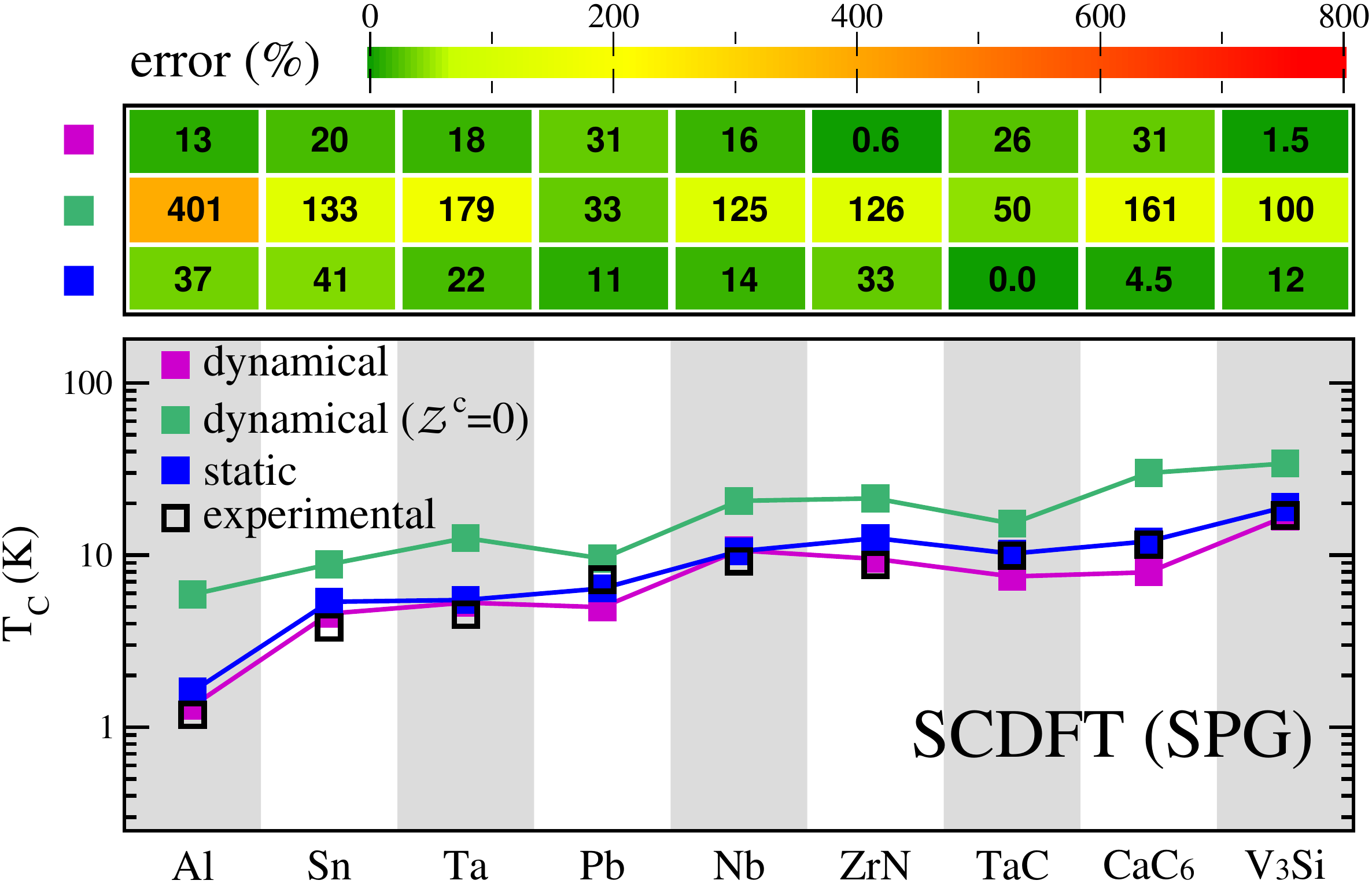}
\caption{Bottom: SCDFT critical temperatures computed using the  LM (left) and SPG (right) phononic functional, within the full dynamical (violet), weak-coupling dynamical (green) and static (blue) approach for the screened Coulomb interaction. Note the use of a logarithmic scale on the ordinate axis. Top: percentage error (computed with respect to the experimental critical temperature).
Errors are highlighted by a color scale from green (accurate) to yellow and red (large deviation).
}\label{fig:tc_comparisonSCDFT}
\end{figure*}


\subsection{SCDFT}\label{sec:SCDFTResults}

In this section we present the results obtained within the SCDFT framework. As for Eliashberg theory, we consider static, plasmonic weak-coupling ($\mathcal{Z}^c=0$, Eq.~(\ref{eq:KcSCDFT}) for $\mathcal{K}^c$) and strong-coupling (Eqs.~(\ref{eq:ZcSCDFT}), (\ref{eq:KcSCDFT})) approaches. These are combined with the treatment of the electron-phonon coupling in the LM~\cite{LuedersSCDFTI2005,MarquesSCDFTII2005} ($\bar{G}_0W$) and SPG~\cite{SPG_EliashbergSCDFT_PRL2020} ($\bar{G}W$) approximations. The calculated \tc's are listed in Tab.~\ref{tab:Tc}, columns D to I, and are compared to the experimental values in Fig.~\ref{fig:tc_comparisonSCDFT}.

By using the phononic LM functional, the SCDFT results obtained within the static approximation for the screened Coulomb interaction underestimate the experimental values by an average error of 35\%. The inclusion of plasmonic effects in both the normal and the superconducting state yields even lower \tc's, with an average error of 43\%. On the other hand, if only the plasmonic contribution to the superconducting pairing is accounted for (i.e., $\mathcal{Z}^c=0$), the theoretical results systematically overestimate the experimental data by a factor of 2. In spite of these deviations, plasmonic SCDFT with the phononic LM functional gives closer \tc's to experiments compared to Eliashberg theory.

As already mentioned, the SPG functional improves over the LM approximation and is comparable in accuracy to conventional Eliashberg theory in describing electron-phonon effects ~\cite{SPG_EliashbergSCDFT_PRL2020}. Employing this functional together with the static Coulomb kernel gives results in good agreement with the experiments. The agreement worsens considerably by including plasmonic effects in the weak-coupling approximation, as this leads to a sizable increase of the predicted \tc's. Nevertheless, adding the plasmonic renormalization mass factor $\mathcal{Z}^c$ suppresses the \tc's values and increases the overall accuracy. The average percentage error in this latter dynamical approach is less than 20\%. For the chosen set of materials, this approximation turns out to be the most accurate, as reported in Tab~\ref{tab:Tc}. However, it should be noticed that all the theoretical results have a non negligible intrinsic error due to the approximations made in calculating the phonon spectral function. For this reason it is not possible to precisely rank in accuracy the different methods. Nevertheless, we can say that plasmonic effects can be safely incorporated in the SCDFT scheme, as they introduce a relatively weak correction to the phonon-induced \tc, which appears to be consistent with the experimental results. 

In Sec.~\ref{sec:EliashbergResults} we have mentioned that the failure of plasmonic Eliashberg theory could be ascribed to the RPA screening or the neglect of Coulomb vertex corrections. The higher accuracy of plasmonic SCDFT, which employs the same Coulomb propagator, indicates that the RPA is not the main source of error. On the other hand, plasmonic SCDFT relies on the $\bar{G}_sW$ approximation for the Coulomb self-energy, whereas Eliashberg theory amounts to the fully self-consistent $\bar{G}W$. This leads us to speculate that vertex corrections to the Coulomb self-energy might be mostly cancelled by the self-consistent dressing of the KS electron Green's function in $\bar{G}_sW$.


\begin{figure}
\includegraphics[width=\columnwidth]{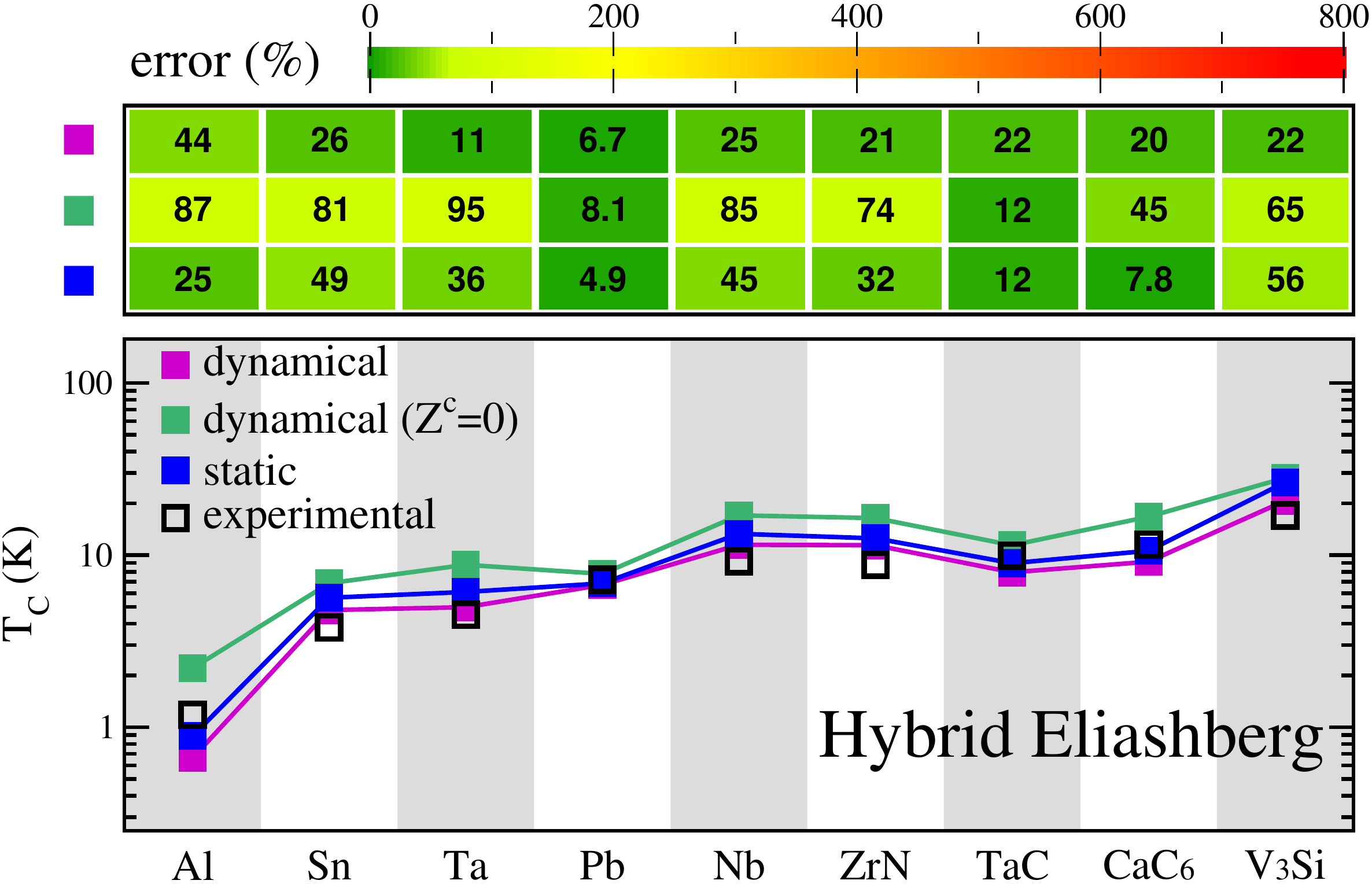}
\caption{Bottom: Hybrid Eliashberg-SCDFT critical temperatures computed within the full dynamical (violet), weak-coupling dynamical (green) and static (blue) approach for the screened Coulomb interaction. Note the use of a logarithmic scale on the ordinate axis. Top: percentage error (computed with respect to the experimental critical temperature).
Errors are highlighted by a color scale from green (accurate) to yellow and red (large deviation).
}\label{fig:tc_comparisonHybrid}
\end{figure}

\begin{table*}\begin{center}
\begin{tabular}{|c||c|c|c||c|c|c||c|c|c||c|c||c|}
\multicolumn{1}{c}~&\multicolumn{1}{c}A&\multicolumn{1}{c}B&\multicolumn{1}{c}C&\multicolumn{1}{c}D&\multicolumn{1}{c}E&\multicolumn{1}{c}F&\multicolumn{1}{c}G&\multicolumn{1}{c}H&\multicolumn{1}{c}I&\multicolumn{1}{c}J&\multicolumn{1}{c}K&\multicolumn{1}{c}L    \\
\hline 
& \multicolumn{3}{c||}{Eliashberg} & \multicolumn{3}{c||}{SCDFT (LM)} & \multicolumn{3}{c||}{SCDFT (SPG)} & \multicolumn{2}{c||}{Hybrid-Eliashberg} & \\
\hline
 & static  & dyn. (Z$^{\rm c}\!=0$)  & dyn. & static & dyn ($\mathcal{Z}^{\rm c}\!=0$) & dyn & static & dyn. ($\mathcal{Z}^{\rm c}\!=0$) &  dyn.  & dyn. (Z$^{\rm c}\!=0$)& dyn. & Exp \\
\hline 
\hline 
Al           &    0.9 &   7.6 &   2.5 &   0.3 &   2.8 &   0.3 &   1.6 &   5.9 &   1.3 &   2.2 &   0.7 &  1.18    \\ \hline                 
Sn           &    5.7 &   9.5 &   6.4 &   3.7 &   7.8 &   3.1 &   5.4 &   8.8 &   4.6 &   6.9 &   4.8 &  3.8     \\ \hline                 
Ta           &    6.1 &  20.3 &  11.0 &   2.8 &   9.6 &   2.9 &   5.5 &  12.6 &   5.3 &   8.8 &   5.0 &  4.5     \\ \hline                 
Pb           &    6.9 &   9.7 &   8.2 &   5.4 &   9.6 &   3.8 &   6.4 &   9.6 &   5.0 &   7.8 &   6.7 &  7.2     \\ \hline                 
Nb           &   13.3 &  41.9 &  23.2 &   7.3 &  19.1 &   7.8 &  10.5 &  20.7 &  10.7 &  17.0 &  11.5 &  9.2     \\ \hline                 
ZrN          &   12.5 &  30.9 &  20.8 &   6.6 &  15.6 &   5.0 &  12.5 &  21.3 &   9.5 &  16.4 &  11.4 &  8.1/9.45\\ \hline                 
TaC          &    9.0 &  19.2 &  13.2 &   4.9 &  10.3 &   3.5 &  10.2 &  15.3 &   7.5 &  11.4 &   7.9 &  9.7/10.2\\ \hline                 
CaC$_{\rm 6}$&   10.6 &  52.3 &  24.8 &   5.9 &  26.0 &   4.2 &  12.0 &  30.0 &   7.9 &  16.7 &   9.2 &  11.5    \\ \hline                 
V$_{\rm 3}$Si&   26.5 & 149.4 &  61.8 &  13.9 &  31.4 &  12.3 &  19.0 &  34.1 &  16.7 &  28.1 &  20.7 &  17      \\ 
\hline 
\hline 
av. \%$|$err$|$& 29.6 &  320.2 &  113.1 &   34.3 &   86.3 &   43.4 &   19.4 &  145.2 &   17.4 &   61.3 &   21.9 &        \\
\hline
av. $|$err$|$&  2.5 &   29.6 &   10.9 &    2.6 &    6.5 &    3.4 &    1.2 &    9.4 &    1.3 &    4.6 &    1.7  &        \\
\hline
max $|$err$|$& 9.5 &  132.4 &   44.8 &    5.6 &   14.5 &    7.3 &    3.1 &   18.5 &    3.6 &   11.1 &    3.7 &    \\
\hline

\end{tabular}\end{center}
\caption{Critical temperatures computed in Eliashberg theory (Secs.~\ref{sec:PlasmonicEliashberg} and~\ref{sec:EliashbergResults}), SCDFT with the phononic LM or SPG functional (Secs.~\ref{sec:SCDFT} and~\ref{sec:SCDFTResults}) and hybrid Eliashberg-SCDFT (Secs.~\ref{sec:hybridEliashbergTheory} and~\ref{sec:HybridResults}), by employing the full dynamical, weak-coupling dynamical and static approach for the screened Coulomb interaction. The corresponding experimental values are also listed. Bottom: percentage and absolute errors. $|{\rm err}|=|$\tc~-~\tcexp~$|$ is the deviation from the experimental \tc. For each method and approximation we indicate the average percentage error av.~\%$|{\rm err}|$=(100\,$|$err$|/$\tcexp), the average error (av. $|{\rm err}|$), and the maximum error (max $|{\rm err}|$) over the material set.
 }\label{tab:Tc}
\end{table*}


\subsection{Hybrid Eliashberg}\label{sec:HybridResults}

From the results of plasmonic Eliashberg theory is evident that the $FW$ approximation for the anomalous Coulomb self-energy significantly overestimates the $T_C$. Since Eliashberg theory is a routinely used method for the prediction of superconducting properties, this appears as a major drawback. A viable alternative is the hybrid Eliashberg-SCDFT approach proposed in  Sec.~\ref{sec:hybridEliashbergTheory}. This, in fact, employs the $F_sW$ approximation, which better describes the plasmonic contribution to the superconducting pairing. 

The \tc's calculated for our test set of materials are collected in Tab.~\ref{tab:Tc} (columns from J to K) and compared to the experimental data in Fig.~\ref{fig:tc_comparisonHybrid}. Consistently with all the previous weak-coupling calculations, the results without the plasmonic mass term tend to overestimate the \tc. In this case the overestimation is, on average, by about 60\%, considerably improving over Eliashberg theory. On the other hand, the fully dynamical approach leads to predicted temperatures that are very close to the experiments.


\section{Conclusions}

We have presented an extension of Eliashberg theory and SCDFT to include the dynamical screening of the Coulomb interaction. Our analysis points at the importance of the plasmonic mass terms, which largely counterbalance the effect of the plasmon-mediated attraction in the Cooper pair. The computational cost associated with the inclusion of the frequency-dependent Coulomb interaction is made affordable by employing an energy-resolved isotropic approximation and by setting nonlinear energy and frequency integration meshes. A hybrid Eliashberg-SCDFT scheme is also formulated, which combines the ME approximation for the electron-phonon coupling with the SCDFT treatment of the dynamically screened Coulomb interaction. The accuracy of the approximations employed in the different methods has been assessed by calculating the plasmon contribution to the critical temperature for a set of classic superconductors. Our simulations show that the SCDFT plasmonic kernels, combined with the phononic SPG functional, yield a good agreement between predicted and measured critical temperatures. Dynamical corrections turn out to be small but not negligible, being of the order of 10-15\% of \tc. Eliashberg theory, although accurate in the static limit of the screened Coulomb interaction, when plasmonic effects are included leads to a large overestimation of \tc, by an average factor of 2. Dynamical Coulomb effects can be included in Eliashberg by adopting the hybrid approach, which gives results close to SCDFT and overall in excellent agreement with experiments.

\section*{Acknowledgments}
We acknowledge financial support by the European Research Council Advanced Grant FACT (ERC-2017-AdG-788890) and German Research Foundation (DFG) through SPP 1840 QUTIF, Grant No. 498/3-1.

\bibliographystyle{apsrev4-1}
\bibliography{paper}
\end{document}